\documentclass[10pt,aps,prb,showpacs,floatfix,floats,superscriptaddress,twocolumn]{revtex4-1}
\usepackage{graphicx,latexsym}
\usepackage{dcolumn}
\usepackage{amssymb,amsmath,bm}

\usepackage{epsfig}

\usepackage[dvips]{color}

\usepackage{hyperref}
\hypersetup{
    pdfnewwindow=true,       
    colorlinks=true,         
    linkcolor=blue,          
    citecolor=blue,          
    filecolor=magenta,       
    urlcolor=black           
}

\def\vec#1{{\bf#1}}
\def\eq#1{Eq.\ (\ref{#1})}
\def\mb#1{\mbox{\boldmath$#1$}}
\def\fig#1{Fig.\ \ref{#1}}

\begin{document}
\title{Single and double finger-gate controlled spin electronic transport\\
 with an in-plane magnetic field}
 \author{Chi-Shung Tang}
 \email{cstang@nuu.edu.tw}
 \affiliation{Department of Mechanical Engineering, National United University,
 Miaoli 36063, Taiwan}
 \author{Jia-An  Keng }
 \affiliation{Department of Mechanical Engineering, National United University,
 Miaoli 36063, Taiwan}
 \author{Nzar Rauf Abdullah}
 \affiliation{Physics Department, College of Science,
University of Sulaimani, Kurdistan Region, Iraq}
 \author{Vidar Gudmundsson}
 \affiliation{Science Institute, University of Iceland,
        Dunhaga 3, IS-107 Reykjavik, Iceland}



\begin{abstract}
A propagation matrix method is proposed to investigate spin-resolved
electronic transport in single finger-gate (SFG) and double
finger-gate (DFG) controlled split-gate quantum devices.  We show
how the interplay of the Rashba and Dresselhaus spin-orbit (SO)
interactions as well as a Zeeman (Z) field influences the quantum
transport characteristics. Without the Dresselhaus effect, the
conductance reveals a mirror symmetry between the hole-like and the
electron-like quasi-bound states in the SO-Z gap energy regime in
the SFG system, but not for the DFG system. For the Dresselhaus
interaction, we are able to analytically identify the binding energy
of the SFG and DFG bound states. Furthermore, the DFG resonant
states can be determined by tuning the distance between the finger
gates.
\end{abstract}

\pacs{73.23.-b, 72.25.Dc, 72.30.+q}


\maketitle

\section{Introduction}

Spintronics utilizing the spin degree of freedom of conduction
electrons is an emerging field of research due to its applications
from logic to storage devices with low power
comsumption.\cite{Loss1998,Zutic2004,Wolf2001} Manipulating the spin
information offers the possibility to scale down devices to the
nanoscale and is favorable for applications in quantum
computing.\cite{Awschalom2002,Awschalom2007,Heedt2012}

Structure inversion asymmetry (SIA) originates from the inversion
asymmetry of the confining potential and yields the Rashba SO
coupling term in the Hamiltonian $H_{\rm R}$, whose strength can be
manipulated by an external field.\cite{Rashba60}  SO interaction
allows for coupling of electron spin and orbital degrees of freedom
without the action of a magnetic field.\cite{Winkler2003,Meier2007}
The Rashba SO coupling is of importance in the study of spintronic
devices in semiconductor materials with two-dimensional electron
gases (2DEG).
\cite{Bandyopadhyay2004,Koo2009,Sadreev2013,Nagaev2014}

Experimentally, the Rashba interaction has been shown to be
effective for electron spin manipulation by using bias-controlled
gate contacts.\cite{Nitta1997} Recently, several approaches were
proposed to engineer a spin-resolved subband structure utilizing
magnetic
fields\cite{Muccio02,Brataas02,Zhang03,Wang03,Serra2005,Scheid2007}
or ferromagnetic materials.\cite{Sun03,Zeng03}  The combination of a
Rashba SO coupling and an external in-plane magnetic field may
modify the subband structure producing a spin-split Rashba-Zeeman
(RZ) subband gap feature.\cite{Pershin2004,Quay2010} To implement a
quantum information storing and transfer, not only coherent
manipulation, but also resonant features involving SO couplings are
of importance.\cite{Zhang2014} This can be achieved utilizing a
double finger gate (DFG) forming a quantum dot in between the
fingers where electrons are subjected to the Rashba SO coupling and
the Zeeman interaction.

Because of the bulk inversion asymmetry (BIA) in III-V
semiconducting materials, the Dresselhaus SO
coupling~\cite{Dresselhaus55} may be induced involving $\bm
k$-linear and $\bm k$-cubic contributions, given by the Hamiltonian
\begin{equation}
  H_{\rm D} =   \beta\left(\sigma_x k_x  - \sigma_y k_y \right) +
              \gamma\left(-\sigma_x k_x k_y^2 + \sigma_y k_y k_x^2
              \right)
\label{Dresselhaus}
\end{equation}
where the strength of the linear in $\bm k$ term $\beta$ = $\gamma
\langle k_z^2 \rangle$ stems from crystal fields. These SO coupling
terms in semiconductor layers are described by the Hamiltonian $
H_{\rm SO}$ = $H_{\rm R}$ + $H_{\rm D}$.

\begin{figure}[tb]
      \includegraphics[width=0.48\textwidth,angle=0]{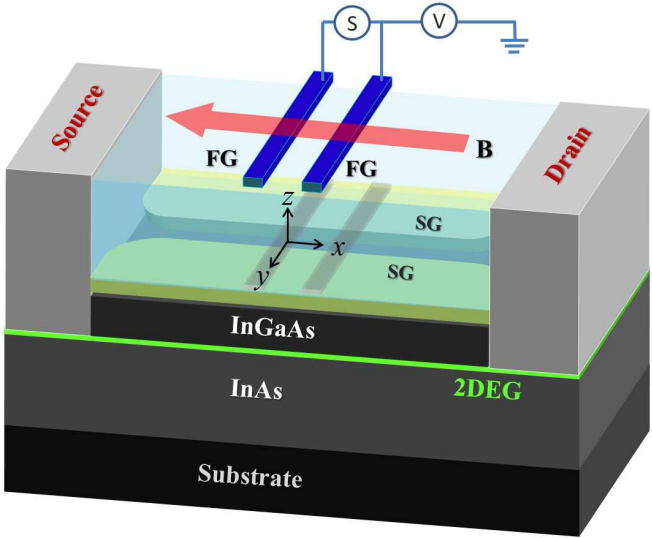}
      \caption{(Color online) Schematic of the SFG and DFG devices controlled by a
switch (S) and a gate potential $V$. A split-gate is used to control
the channel width. An external in-plane magnetic field $\vec{B} = B
\hat{\mb{x}}$ ($B<0$) is applied. The DFG is consisted of two finger
gates located $x_1$ and $x_2$ to influence the spin-resolved
resonant quantum transport.} \label{fig1}
\end{figure}

In this work, we consider a finger-gate (FG) controlled narrow
constriction\cite{Tang2012,Loss2014} in the presence of a RZ subband
gap, in which a very asymmetric structure in the 2DEG leads to
strong SO coupling with the result that the Rashba effect is
dominant. The Dresselhaus effect due to BIA is also considered.
Below, the SFG and DFG controlled spin-resolved electronic transport
properties will be compared in an external in-plane magnetic field
as shown in \fig{fig1}.

The organization of the rest of this paper is as follows. In Sec.\
II we describe the propagation-matrix approach of tunneling through
a DFG system under in-plane magnetic field. In Sec.\ III we present
our calculated results on the spin-split subband structure and
the spin-resolved conductance. A concluding remarks is given in Sec.\
IV.


\section{ spin electronic transport model}

In this section, we show that the RZ and RD-Z effects, described by
a Hamiltonian technique lead to, respectively, symmetric and
asymmetric spin-split subband structures. In addition, the SFG and
DFG influence on the transport through the SG confined quantum
channel will be described by a spin-dependent PM method.

As is illustrated for the device in \fig{fig1}, a two dimensional
electron gas (2DEG) is induced in an InAs-In$_{1-x}$Ga$_x$As
semiconductor heterojunction grown in the $(001)$ crystallographic
direction and is subjected to a split-gate voltage. A pair of
split-gates restricts the movement of the electrons of the 2DEG, and
therefore a quantum channel is generated in the $[100]$ direction.
Propagating electrons in the channel are driven from source to
drain.

In the absence of the finger gates, a transported electron is
affected by the Rashba effect $H_R$ due to SIA and the Zeeman effect
$H_Z$ induced by an external in-plane magnetic field, described by
the unperturbed Hamiltonian
\begin{equation}
\widetilde{H}_0 = H_0 + H_{\rm R}+ H_{\rm D} + H_{\rm Z}.
 \label{H0til}
\end{equation}
The first term describes a bare quantum channel that is described by
the ideal Hamiltonian
\begin{equation}
 H_0 = \frac{{\hbar ^2}{k^2}}{{2{m^*}}} + {U_{\rm SG}}(y).
\end{equation}
The first term is the kinetic energy of an electron in the 2DEG,
where $\hbar = h/2\pi$ is the reduced Planck constant. A conduction
electron has an assigned wave number $k$ satisfying $k^2=k_x^2 +
k_y^2$ and $m^*$ is its effective mass.  The second term indicates a
split-gate induced confining potential energy that can be modeled by
a hard-wall confinement with width $W$ for simplicity.

In the second term of \eq{H0til}, we consider a $(001)$
crystallographic 2DEG system, and hence the Rashba SO Hamiltonian
$H_{\rm R} = \alpha \left( \mb{\sigma}\times \mathbf{k}\right) \cdot
\hat{\mathbf{z}}$ coupling the Pauli spin matrix $\mb{\sigma}$ to
the momentum $\mathbf{p}= \hbar \mathbf{k}$ can be reduced to a
$k$-linear form
\begin{equation}
H_{\rm R} =  \alpha \left( \sigma_x k_y - \sigma_y k_x \right),
\label{HR}
\end{equation}
\begin{equation}
{H_{\text{D}}} = \beta ({\sigma _x}{k_x} - {\sigma _y}{k_y}),
\label{HD}
\end{equation}
where the Rashba coupling strength $\alpha$ is proportional to the
electric field along the $\hat{\mathbf{z}}$ direction perpendicular
to the 2DEG.\cite{Nitta1997} The fourth term in \eq{H0til} describes
an applied external in-plane magnetic field that is selected to be
antiparallel to the channel in the $[100]$ direction and has the
form $\vec{B} = B \hat{\mathbf{x}}$ ($B<0$). The longitudinal
in-plane magnetic field induces a Zeeman term that can be expressed
as
\begin{equation}
H_Z = g{\mu_B}B \sigma_x,
\end{equation}
in which $g = g_s/2$ indicates half of the effective gyromagnetic
factor ($g_s = -15$ for InAs) and $\mu_B = 5.788\times
10^{-2}$~${\rm meV/T}$ is the Bohr magneton.

Tuning the switch shown in \fig{fig1} allows us to investigate SFG
and DFG controlled spin-resolved transport properties. We consider
the width of the finger-gate scattering potential, $W$, to be less
than the Fermi wave length $\lambda_F = 31.4$~nm and to be described
by a delta potential. In addition, we assume a high-mobility
semiconductor materials so that impurity effects can be neglected. A
FG array system can generally be described by the scattering
potential energy
\begin{equation}
 U_{\rm FG}(x) =  \sum_{j=1}^{N_{\rm FG}} U_j \delta(x-x_j),
 \label{Usc}
\end{equation}
where $U_j$ indicates a delta potential energy induced by the FG
$j$, and $N_{\rm FG}$ is the number of FG. For example,  $N_{\rm
FG}$ = 1 and 2 indicate, respectively, SFG or DFG systems. These FG
array systems under the influence of the RD-Z effects can be
formally described by the Schr\"{o}dinger equation
\begin{equation}
\left[ \widetilde{H}_0 + U_{\rm FG}(x) \right] \Psi (x,y) = E \Psi
(x,y). \label{Htotal}
\end{equation}
The eigenfunction $\Psi (x,y)$ in \eq{Htotal} can be obtained by
summing over all occupied subbands, $n$, for the product of the
spatial wave functions and the spin states, given by
\begin{equation}
 \Psi (x,y) = \sum_n {\phi _n}(y){e^{i{k_x}x}}\chi_n \, .
 \label{Psi}
\end{equation}
Here the ideal transverse wave function in subband $n$ is $\phi_n(y)
= (2/W)^{1/2} \sin(k_n y)$,  in which the subband energy
 $\varepsilon_{n} = \hbar^2 k_n^2 / 2m^*$ with the quantized wave number $k_n = n\pi/W$.

For simplicity, we employ the Fermi-level in a 2DEG as an energy
unit, namely $E^*$ = $E_F$ = $\hbar^2 k_F^2/2m^*$ with $m^\ast$ and
$\hbar$ being, respectively, the effective mass of an electron and
the reduced Planck constant. In addition, one selects the inverse
wave number as a length unit, namely  $l^* = k_F^{-1}$.
Correspondingly, the magnetic field is in units of $B^* = \mu_B^{-1}
E^*$, and the Rashba SO-coupling constant $\alpha$ is in units of
${\alpha}^* = 2E^*l^*$.  In the following we consider a sufficient
narrow channel by assuming the channel width $W=\pi l^*=15.7$~nm so
that the bare subband energy is simply $\varepsilon_{n} = n^2$. The
energy dispersion can thus be expressed as
\begin{equation}
{E_n^\sigma ({k_x}) = {\varepsilon _{n}} + k_x^2 + \sigma \sqrt
{{{(2\beta {k_x} + gB)}^2} + {{(2\alpha {k_x})}^2}}},
 \label{En_kx}
\end{equation}
where $\sigma = \pm$ indicates the upper ($+$) and lower ($-$) spin
branches.  Sufficiently low temperature $k_B T < 0.1 \Delta
\varepsilon$ or $T < 23$~K is required to avoid thermal broadening
effects.


In order to investigate the SFG and DFG controlled spin-resolved
electronic transport properties, we shall explore how the
spin-mixing effect due to the interplay of the RD-Z effects
influences the propagating and evanescent modes.  For a given
incident electron energy $E_n = E - \varepsilon_n$ in the subband
$n$, the energy dispersion is related to complex wave number that
obeys
\begin{eqnarray}
&& k_x^4 - \left[ {4({\alpha ^2} + {\beta ^2}) + 2E_n} \right]k_x^2
- {\text{4}}gB\beta {k_x} \nonumber \\
 &&+ \left[ {E_n^2 - {{(gB)}^2}} \right] = 0\, .
 \label{kx_En}
\end{eqnarray}
 To proceed, one has to label the four longitudinal
wave numbers $k_x$ as the right-going $k^{\sigma}$ and left-going
$q^{\sigma}$, in which the notation $\sigma = +$ indicates spin-up
mode and $\sigma = -$ stands for spin-down mode.

Below, we focus on a sufficiently narrow quantum channel to explore
the first two conductance steps associated with the two spin
branches of a transported electron occupying the lowest subband.  We
calculate the quantum transport properties by using a generalized PM
method, in which the spin-flip scattering mechanisms is taken into
account. The energy dispersion shown in \fig{fig2}(a) essentially
divides the energy spectrum into three regimes, namely the low
energy regime $E^-_{0{\rm R}} < E < E^-_{0{\rm T}}$, the
intermediate energy regime $E^-_{0{\rm T}} < E < E^+_{0}$, and the
high energy regime $E > E^+_{0}$.   In the low and high energy
regimes, there are four propagating modes with real $k_{\sigma}$ and
real $q_{\sigma}$.  It should be noted that there are two
propagating and two evanescent modes in the intermediate energy
regime or the RZ energy gap region where the evanescent modes
manifest a bubble behavior with imaginary wave
vectors.\cite{Tang2012}

The spin-split wave functions around the scattering potential
$U_{\rm FG}$ located at $x_j$ can be formally expressed including
the spatial and spin parts as
\begin{eqnarray}
 \mb{\psi}_{j - 1}\left( x \right) &&= \sum\limits_{\sigma  =  \pm }
 \sum\limits_{\sigma^\prime  =  \sigma, \bar{\sigma} }
 \left\{
{A_{j - 1}^{\sigma,\sigma^\prime} {e^{ik^{\sigma^\prime} \left( {x -
{x_j}} \right)}} \left[ {\begin{array}{*{20}{c}}
{a^{\sigma^\prime} }\\
{b^{\sigma^\prime} }
\end{array}} \right]}
\right.
\nonumber \\
 +&&
 \left.
 {B_{j - 1}^{\sigma,\sigma^\prime} {e^{iq^{\sigma^\prime} \left( {x - {x_j}} \right)}}\left[ {\begin{array}{*{20}{c}}
{c^{\sigma^\prime} }\\
{d^{\sigma^\prime} }
\end{array}} \right]} \right\}, \quad x <
x_j \label{wf1}
\end{eqnarray}
and
\begin{eqnarray}
 \mb{\psi} _{j}\left( x \right) &&= \sum\limits_{\sigma  =  \pm }
 \sum\limits_{\sigma^\prime  =  \sigma, \bar{\sigma} }
 \left\{
{C_{j}^{\sigma,\sigma^\prime} {e^{ik^{\sigma^\prime} \left( {x -
{x_j}} \right)}}\left[ {\begin{array}{*{20}{c}}
{a^{\sigma^\prime} }\\
{b^{\sigma^\prime} }
\end{array}} \right]}\right. \nonumber \\
 +&& \left.
 {D_{j}^{\sigma,\sigma^\prime} {e^{iq^{\sigma^\prime} \left( {x - {x_j}} \right)}}\left[ {\begin{array}{*{20}{c}}
{c^{\sigma^\prime} }\\
{d^{\sigma^\prime} }
\end{array}} \right]} \right\}, \quad x >
x_j \label{wf2}
\end{eqnarray}
where $A_{j-1}^{\sigma,\sigma^\prime}$ and
$C_{j}^{\sigma,\sigma^\prime}$ indicate the coefficients of
right-going electrons with positive group velocity and wave number
$k^\sigma$, while $B_{j-1}^{\sigma,\sigma^\prime}$ and
$D_j^{\sigma,\sigma^\prime}$ stand for the coefficients of the
left-going electrons with negative group velocity and wave number
$q^\sigma$. There are two boundary conditions around $x_j$, given by
\begin{subequations}
\begin{align}
        \mb{\psi}_{j-1} \left( {x_j^- } \right) &= \mb{\psi}_{j} \left(
{x_j^+} \right) , \label{bc1} \\
        \mb{\psi}_{j-1} '\left( {x_j^-} \right) &= \mb{\psi}_{j} '\left(
{x_j^+} \right) - {U_j}\mb{\psi}_{j} \left( {x_j^+} \right) .
\label{bc2}
\end{align}
\end{subequations}

Taking into account the possible incident spin states $\sigma$ and
$\bar{\sigma }$ allows us to formulate the total PM $\rm{\bf P}$ in
an arbitrary FG array system.
\begin{equation}
\left[ {\begin{array}{*{20}{c}}
   {\bf 1}  \\
   {\bf r}  \\
\end{array}} \right]
 = {{\bf{P}}}
 \left[ {\begin{array}{*{20}{c}}
   {\bf t}  \\
   {\bf 0} \\
\end{array}} \right]\, ,
 \label{TPM}
\end{equation}

or expressed explicitly
\begin{equation}
\left[ {\begin{array}{*{20}{c}}
   1 & 0  \\
   0 & 1  \\
   {r_{\sigma ,\sigma }} & {r_{\bar \sigma ,\sigma }}  \\
   {r_{\sigma ,\bar \sigma }} & {r_{\bar \sigma ,\bar \sigma }}  \\
\end{array}} \right]
 = {\bf{P}}
 \left[ {\begin{array}{*{20}{c}}
   {t_{\sigma ,\sigma }} & {t_{\bar \sigma ,\sigma }}  \\
   {t_{\sigma ,\bar \sigma }} & {t_{\bar \sigma ,\bar \sigma }}  \\
   0 & 0  \\
   0 & 0  \\
\end{array}} \right]\, .
 \label{PME}
\end{equation}
Here, the diagonal and off-diagonal terms in  $\mathbf{r}$ and
$\mathbf{t}$ indicate, respectively, the spin-preserve (SP) and
spin-flip (SF) reflection and transmission coefficients.  The first
subscript is the incident spin state, and the second one is the
scattered spin state.

To proceed, one has to consider both the $\sigma$ and $\bar{\sigma
}$ spin states incident from the source electrode.  The PM for the
electrons with two spin states in the $j$-th region can be expressed
as $\mathbf{p}_j = \mathbf{p}_{j,\delta} \mathbf{p}_{j,{\rm free}}$,
in which $\mathbf{p}_{j,\delta}$ is the PM through the FG $j$ and
$\mathbf{p}_{j,{\rm free}}$ is the free space PM between the FG $j$
and $j+1$.  Hence, the total PM for SFG and DFG can be simply
expressed as $\mathbf{P}_{\rm SFG} = \mathbf{p}_1$ and
$\mathbf{P}_{\rm DFG} = \mathbf{p}_1\mathbf{p}_2$.  In addition,
$\mathbf{p}_{j,{\rm free}}(i,j) = \exp(-ik_{i,j}L)$ if $i=j$, or
identically zero if $i \ne j$,
 in which $k_{1,1}=k^{\sigma}$ and $k_{2,2}= k^{\bar \sigma}$ are right-going wave vectors, while $k_{3,3}=
q^{\sigma}$ and $k_{4,4}= q^{\bar \sigma}$ are left-going wave
numbers. Solving the PM equation numerically, we may obtain the
reflection and transmission coefficients of the scattered
intermediate and final states through the SFG or DFG systems.

We consider an electron injected from the left reservoir (source
electrode) and transported to the right reservoir (drain electrode)
for a given incident energy.  Solving for the spin non-flip and flip
reflection coefficients $r_{\sigma ,\sigma}$ and $r_{\sigma ,\bar
\sigma}$, as well as the spin non-flip and flip transmission
coefficients $t_{\sigma ,\sigma}$ and $t_{\sigma ,\bar \sigma}$, we
can calculate numerically the conductance based on the
Landauer-B\"{u}ttiker framework\cite{Landauer1970,Buttiker1990}
\begin{equation}
G = g_0 \sum\limits_{\sigma_L,\sigma_R}
 \frac{v_{\sigma_R}}{v_{\sigma_L}}
 \left| t_{\sigma_L,\sigma_R} \right|^2 \, .
\label{eq3.2.30}
\end{equation}
Here $g_0$ = $e^2/h = 25.8$~${\rm k}\Omega^{-1}$ is the conductance
quantum per spin state, and $\sigma_L$ and $\sigma_R$ indicate,
respectively, the spin branches of the incident and transmitted
waves in the left and right leads. Therefore, ${v_{{\sigma_L}}}$ and
${v_{{\sigma_R}}}$ represent the group velocity of corresponding
modes in the left and right reservoirs, respectively.


\section{Numerical Results}

Calculations presented below are carried out under the assumption of
a 2DEG at a high-mobility InAs-In$_{1-x}$Ga$_x$As semiconductor
interface with an electron effective mass $m^{\ast}=0.023 m_0$ and
typical electron density $n_e \sim
10^{12}$~cm$^{-2}$.\cite{Nitta1997} Accordingly, the energy unit is
$E^*$ = 66~meV, the length unit $l^*$ = 5.0~nm, the magnetic field
unit $B^* = 1.14$~$\textrm{kT}$, and the spin-orbit coupling
parameter is in units of $\alpha^*$ = 330~meV$\cdot$nm.  In
addition, we assume that the width of the finger gate is $l_{\rm
FG}$ = $l^*$ such that the FG potential energy $V_j$ = $U_j/l_{\rm
FG}$ is in units of $V^* = 66$~meV. Below, we assume that $V_1$ =
$V_2$ = $V$ for simplicity.  By using the above units, all physical
quantities presented below are dimensionless.\cite{Tang2012}

\begin{figure}[tb]
\begin{center}
\mbox{
 {\includegraphics[width=0.2\textwidth]{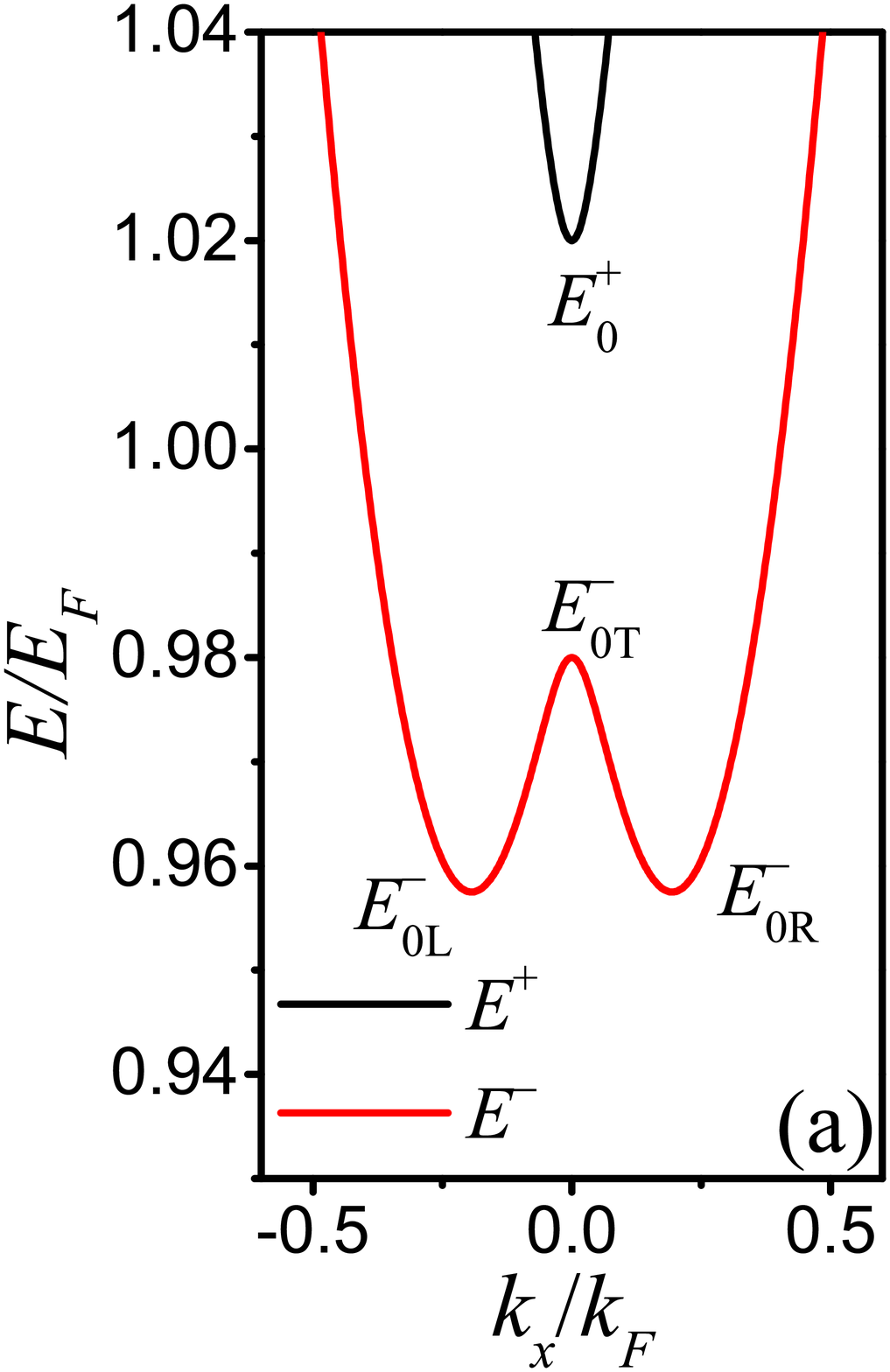}}
 {\includegraphics[width=0.2\textwidth]{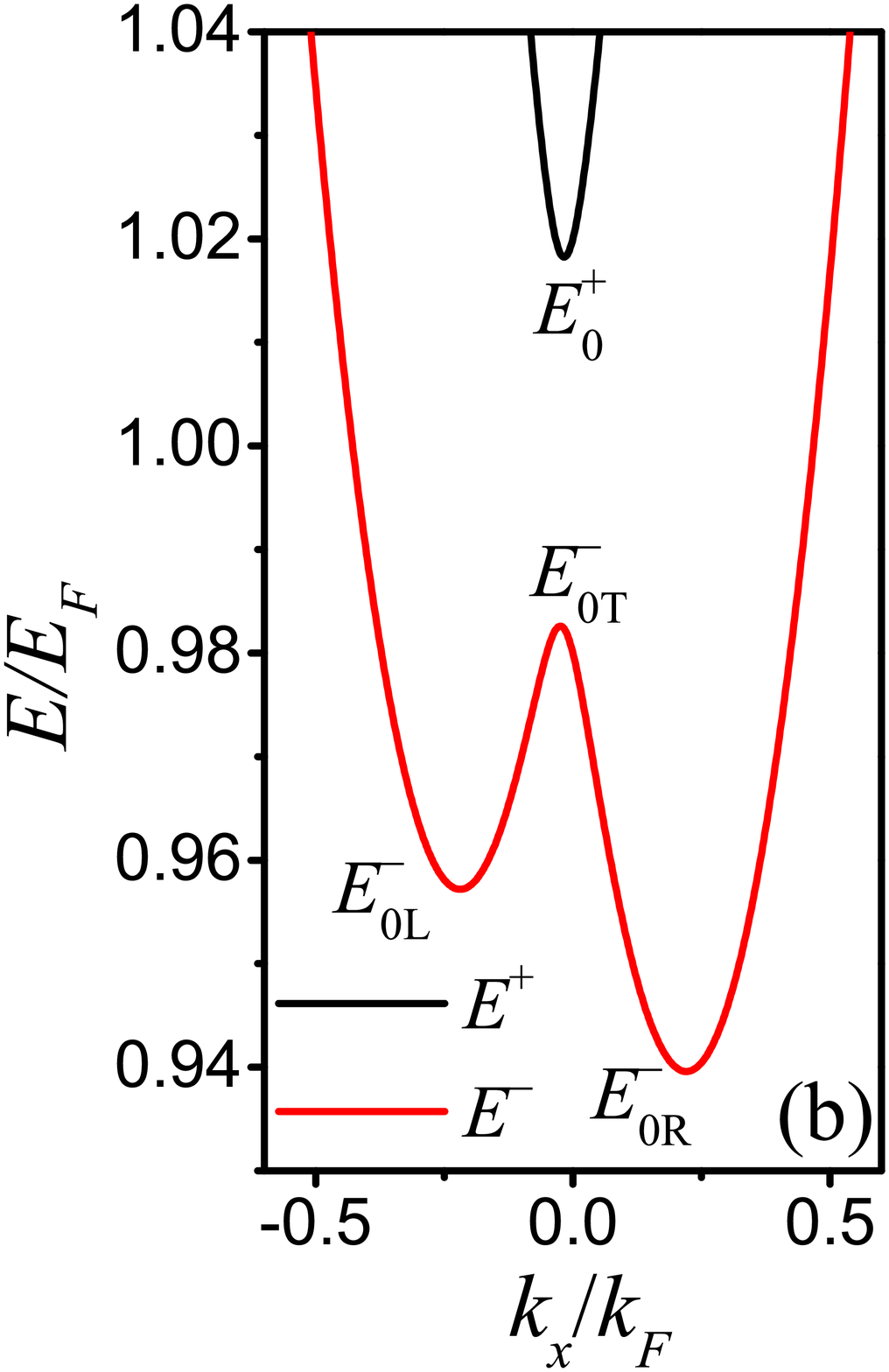}}
 }
 \mbox{
 {\includegraphics[width=0.22\textwidth]{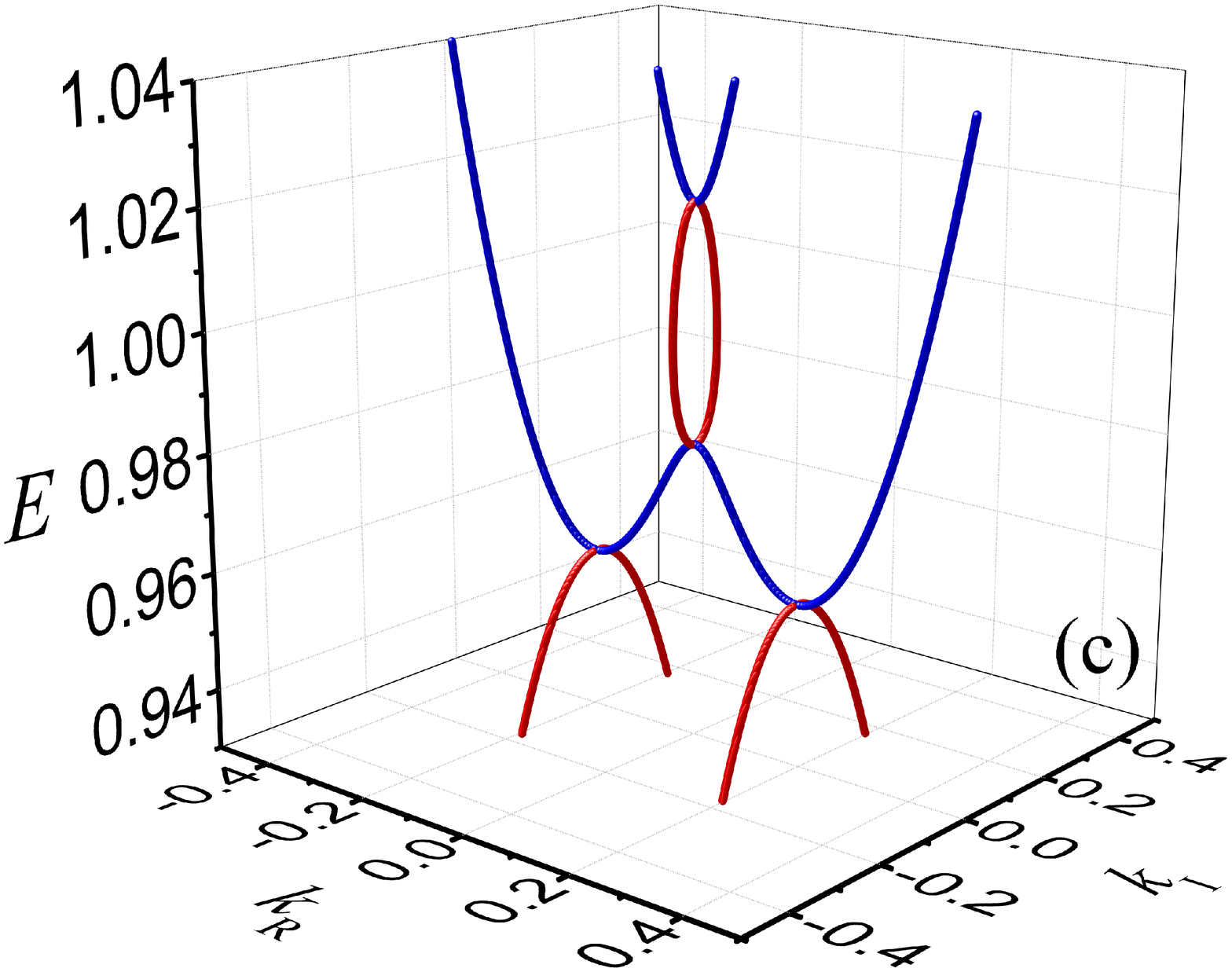}}
 {\includegraphics[width=0.22\textwidth]{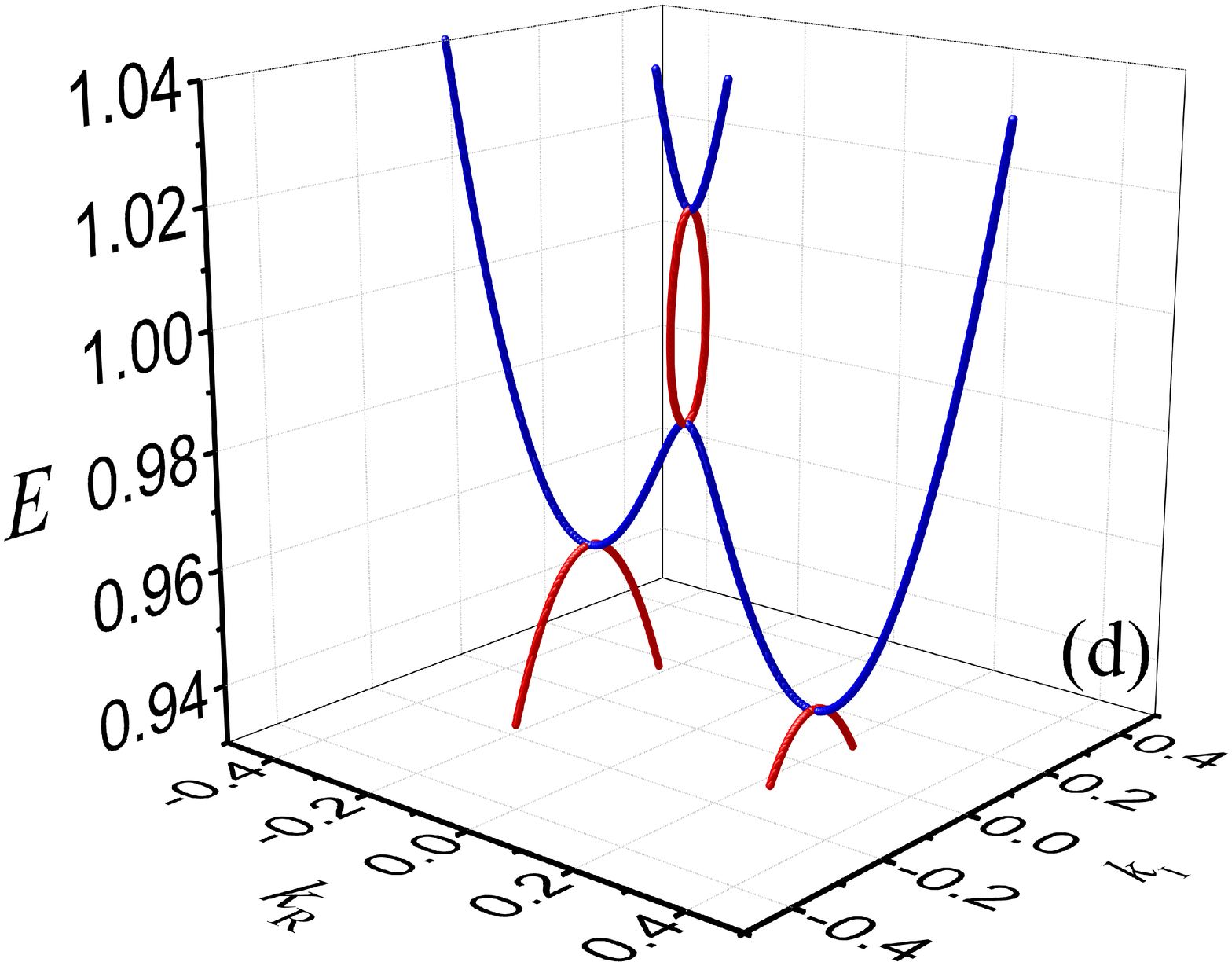}}
 }
\end{center}
\caption{(Color online) (a)-(b) Energy spectrum as a function of
real wave number;  (c)-(d) propagating (blue) and evanescent (red)
modes versus complex wave number $k = k_R + i k_I$. The Dresselhaus
coefficients are $\beta = 0.0$ for (a) and (c); $\beta = 0.1$ for
(b) and (d). The other parameters are $\alpha = 0.2$, $gB = 0.02$.}
 \label{fig2}
\end{figure}

Our previous work has demonstrated that the interplay of the SO
interaction and the Zeeman effect may generate a SO-Z gap due to the
orthogonality of the SO effective magnetic field and the in-plane
magnetic field.\cite{Tang2015}  In \fig{fig2}, we show the energy
spectrum of the lowest subband.  The chosen parameters correspond to
a strong SO coupling regime with SO-Z gap.  The transported
spin-resolved electrons can thus be separated into the low-energy
($E_{0{\rm R}}^- <E <E_{0{\rm T}}^-$), the SO-gap ($E_{0{\rm T}}^- <
E < E_{0{\rm T}}^+$), and the high-energy ($E_{0}^+ < E$) regimes.
Figure \ref{fig2}(a)-(b) indicates two spin-state energies for a
given wave number obtained from \eq{En_kx}, while Figure
\ref{fig2}(c)-(d) displays four complex wave numbers for a given
electron energy obtained from \eq{kx_En}.

Concerning the Rashba-Zeeman (RZ) effect, \fig{fig2}(a) shows that
the subband bottom energy of the spin-up branch is at $E_{0}^+$ =
$\varepsilon_1 + gB$ = 1.02, and the  subband top energy of the
spin-down branch is $E_{0{\rm T}}^-$ = $\varepsilon_1 - gB$ = 0.98.
Hence, the RZ gap $\Delta {E_{\rm{RZ}}}$ = $E_{0}^+ - E_{0{\rm
T}}^-$ = $2gB$ = 0.04.  This is exactly the Zeeman gap $\Delta
{E_{\rm{Z}}}$.  In addition, the left and right spin-down subband
bottoms are at the same energy, namely $E_{0{\rm L}}^- = E_{0{\rm
R}}^- = 0.9575$.

Concerning the Rashba-Dresselhaus-Zeeman (RDZ) effect ($\beta =
0.1$),  \fig{fig2}(b) shows that spin-up subband bottom becomes
slightly lower ($E_{0}^+$ = 1.018), and the spin-down subband top
becomes slightly higher ($E_{0{\rm T}}^-$ = 0.982). Hence, the RDZ
gap $\Delta {E_{\rm{RDZ}}}$ = 0.036 is smaller than the RZ gap
$\Delta {E_{\rm{RZ}}}$ by 0.004.  Moreover, the left and right
spin-down subband bottoms are no longer the same, that is, $E_{0{\rm
L}}^- = 0.9571$ and $E_{0{\rm R}}^- = 0.9395$. Below, we shall show
that these asymmetric subband bottoms may lead to interesting
transport properties.

In order to explore the spin-resolved transport properties, it is
important to define the group velocity of an electron in the
$\sigma$ spin branch
\begin{equation}
{v^\sigma }({k_x})  = 2{k_x} + \sigma \frac{{4({\alpha ^2} + {\beta
^2}){k_x} + 2\beta gB}}{{\sqrt {{{(2\alpha {k_x})}^2} + {{(2\beta
{k_x} + gB)}^2}} }}\, .
 \label{vg}
\end{equation}
Defining the velocity allows us to determine a local minimum and a
maximum in the subband structures by setting the group velocity
identically zero.

Without the Dresselhaus effect, the two subband bottoms in the lower
spin branch can be analytically obtained at $k_x = \pm \left[
\alpha^2 - (gB/2\alpha)^2 \right]^{1/2}$ corresponding to the same
subband bottom energy $E_{0{\rm L}}^-$ = $E_{0{\rm R}}^-$ = $1 -
\left[ \alpha^2 +(gB/2\alpha)^2\right]$. With the Dresselhaus
effect, this degeneracy subband bottom will be broken as is shown in
\fig{fig2}(b). \fig{fig2}(c) and (d) show their corresponding energy
dispersion with respect to complex $k_x$ that is important to
perform transport calculation.

\begin{figure}[t]
\includegraphics[width = 0.48 \textwidth, angle=0] {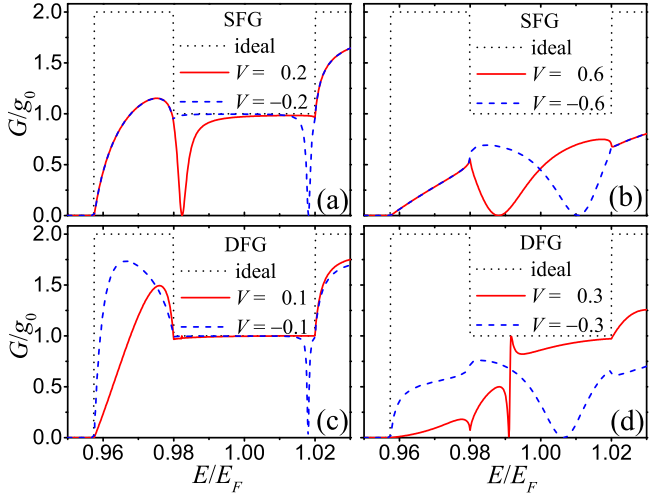}
\caption{(Color online) Conductance as a function of energy in the
absence of Dresselhaus effect for (a)-(b) SFG system, in comparison
with (c)-(d) DFG system with $L = 10$. $\alpha = 0.2$, $\beta =
0.0$, and $gB = 0.02$.}\label{fig3}
\end{figure}

In \fig{fig3}, we compare the conductance behavior of SFG and DFG
systems with no Dresselhaus effect.  The conductance in the SFG
system shown in \fig{fig3} (a) and (b) reveals a perfect mirror
effect between the electron-like QBS (EQBS) for negative $V$ and
hole-like QBS (HQBS) for positive $V$. Moreover, the QBS dips for $V
= \pm 0.2$ become the QBS valleys for $V = \pm 0.6$. This indicates
that a stronger voltage results in a shorter QBS life time.

It is interesting to compare the SFG with voltage $V$ to a DFG with
voltage $V/2$ and keep the two FGs very close, say with $L =10$, as
shown in \fig{fig3} (c) and (d). It is surprising that the EQBS
structures in the DFG and the SFG system at a negative $V$ are
almost unchanged, but the HQBS structures are very different.
\fig{fig3}(c) demonstrates that a weak positive voltage ($V=0.1$) in
the DFG system does not allow a formation of a HQBS.  If the voltage
is increased to 0.3, the HQBS valley in SFG becomes a dip around
$E_{0T}^-$ and a Fano dip in the SO-Z gap as shown in \fig{fig3}(d).

\begin{figure}[t]
\includegraphics[width = 0.48 \textwidth,angle=0] {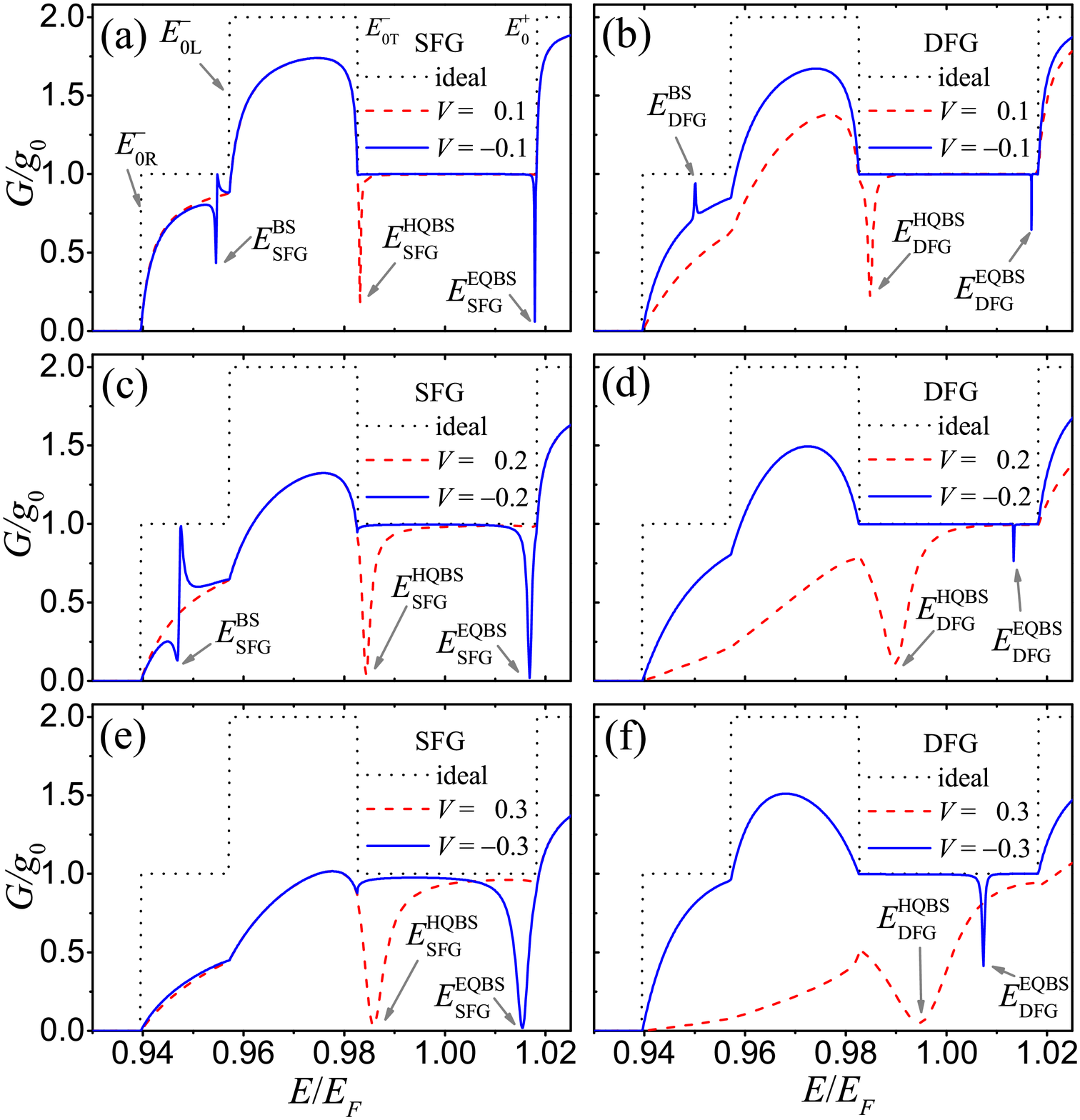}
\caption{(Color online) Conductance as a function of energy in the
presence of Dresselhaus effect for SFG system in comparison with DFG
system with $L = 5$. $\alpha = 0.2$, $\beta = 0.1$, and $gB =
0.02$}.
 \label{fig4}
\end{figure}

In \fig{fig4}, we show the conductance of SFG and DFG systems as a
function of electronic energy in the presence of Dresselhaus SO
interaction $(\beta = 0.1)$ for $V$ = (a)-(b) $\pm 0.1$; (c)-(d)
$\pm 0.2$; and (e)-(f) $\pm 0.3$. The other parameters $\alpha =
0.2$ and $gB = 0.02$, satisfying $\alpha^2$ + $\beta^2 > gB$, are
within the strong SO coupling regime.

In the case of a SFG, sufficient low energy $E < E_{0L}^-$ with
negative FG voltage may result in a Fano line-shape (see
\fig{fig4}(a) and (c)).  This is due to the interference between the
SFG BS $E_{\rm SFG}^{\rm BS}$ below the left spin-down bottom and
the extended state at the right spin-down branch.  The binding
energy of the SFG BS can be analytically predicted to be $E_{\rm
SFG}^b$ = $V^2/4$ = 0.0025 ($V=-0.1$) and 0.01 ($V=-0.2$), as shown
by solid blue lines. This binding energy can also be numerically
determined using
\begin{equation}
E_{\rm SFG}^b  = E_{0{\rm L}}^- - E_{\rm SFG}^{\rm BS} \,
\end{equation}
with $E_{0{\rm L}}^-$ = 0.957, while $E_{\rm SFG}^{\rm BS}$ = 0.9545
($V=-0.1$) and 0.9468 ($V=-0.2$). We thus numerically obtain the SFG
BS binding energy $E_{\rm SFG}^b$ = 0.0026 ($V=-0.1$) and 0.0103
($V=-0.2$) that is approximately the same as our analytical
prediction delivers.

If the incident electron energy is within the gap region, the mirror
effect between the HQBS and EQBS is clearly shown in the SFG system.
More precisely, the SFG HQBS energies $E_{\rm SFG}^{\rm HQBS}$ =
0.9831, 0.9843, 0.9857 for $V$ = 0.1, 0.2, 0.3, respectively, are
slightly above the spin-down top energy $E_{0{\rm T}}^-$ = 0.982.
Accordingly, the SFG EQBS energies $E_{\rm SFG}^{\rm EQBS}$ =
1.0178, 1.0168, 1.0153 for $V$ = $-0.1, -0.2, -0.3$, respectively,
are slightly below the spin-up bottom energy $E_{0}^+$ = 1.018. The
higher $|V|$ may slightly shift the HQBS and EQBS toward the center
of the SO-Z gap.

In the case of a DFG shown in \fig{fig4}, a sufficient low energy $E
< E_{0{\rm L}}^-$ = 0.957 with a negative FG potential $V = -0.1$
may result in a sharp peak (see \fig{fig4}(b)) corresponding to a
DFG BS energy $E_{\rm DFG}^{\rm BS}$ = 0.9501 below $E_{0{\rm
L}}^-$. The DFG BS binding energy can be numerically obtained
\begin{equation}
E_{\rm DFG}^b  = E_{0{\rm L}}^- - E_{\rm DFG}^{\rm BS}
\end{equation}
giving $E_{\rm DFG}^b$ = 0.0069 in units of $E_F$.

In order to provide an evidence of such a DFG BS mechanism, we
derive an analytical expression for $E_{\rm DFG}^b$, given by
\begin{equation}
\frac{|V|}{\sqrt{E_{\rm DFG}^b}}-1
 = \tanh \left( \sqrt{E_{\rm DFG}^b} \frac{L}{2} \right) \, .
 \label{Eb_DFG}
\end{equation}
This equation allows us to analytically estimate the binding energy
$E_{\rm DFG}^b$ = $0.00689$ for $V = - 0.1$ and $L=5$.  This is only
a bit smaller than our numerical result.

Comparing Figs.\ \ref{fig4}(b), (d), and (f), we see that if the DFG
potential is negatively increased, the electron-like QBS (EQBS) dip
is red shifted towards the center of the SO-Z gap and becomes more
significant. If $V = - 0.2$, as is shown in \fig{fig4}(d), the DFG
binding energy $E_{\rm DFG}^b = 0.02$ as can be estimated from
\eq{Eb_DFG}, the corresponding BS energy $E_{\rm DFG}^{\rm BS} =
0.937$ is below the subband threshold $E_{0{\rm R}}^-$.  The DFG BS
thus disappears in \fig{fig4}(d) as well as \fig{fig4}(f).  On the
contrary, increasing the positive DFG potential ($V>0$), the HQBS
dip around the $E^-$ subband top becomes a blue shifted broad
valley, as shown by the dashed red lines in Figs. \ref{fig4}(b), (d)
and \ref{fig4}(f). These HQBS dips are at $E = 0.985,$ $0.990,$ and
$0.995$, respectively. The shift of the location of the HQBS is
$\delta E = 0.005$ if the DFG potential is increased by $\delta V =
0.1$.

\begin{figure}[t]
\includegraphics[width = 0.44 \textwidth,angle=0] {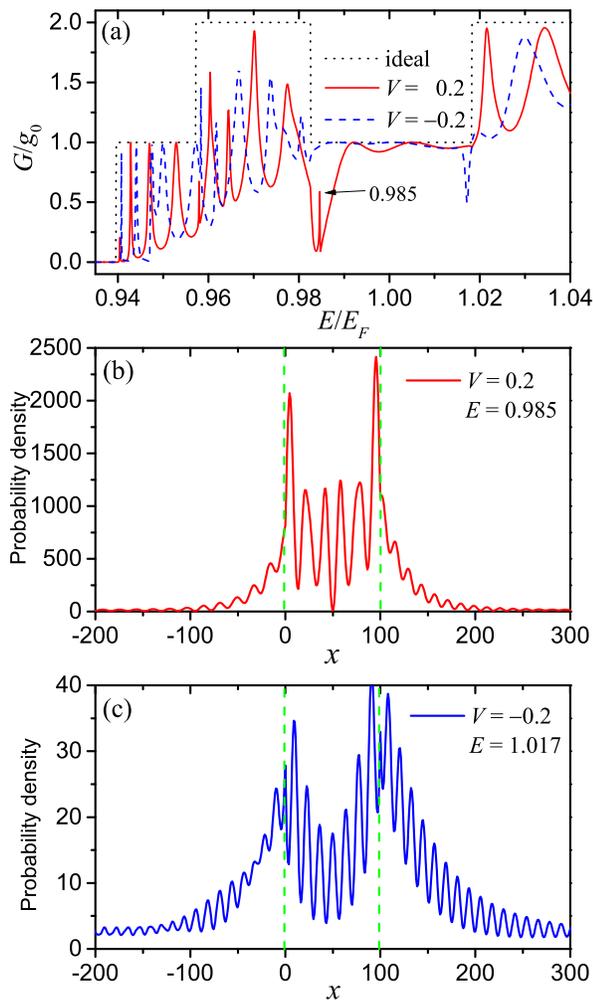}
\caption{(Color online) (a) Conductance as a function of energy for
DFG system with $L = 100$ (500 nm). The corresponding probability
densities are shown in (b) $V$ = $0.2$, $E$ = $0.985$; and (c) $V$ =
$-0.2$, $E$ = $1.017$. Other parameters are $\alpha = 0.2$, $\beta =
0.1$, and $gB = 0.02$}.
 \label{fig5}
\end{figure}

In order to demonstrate the possibility of forming a resonant state
(RS) in a DFG system, we consider the system with a long FG
distance, $L$ = $100$. The corresponding conductance is plotted as a
function of energy shown in \fig{fig5}(a). When the gate voltage $V$
is $-0.2$ (short dashed line), the conductance manifests a clear
EQBS resonance dip at energy $E$ = $1.017$. However, for a positive
gate voltage $V=0.2$ (solid line) the conductance displays a more
complicated structure. There is a broad resonance leading to a
valley structure around $E$ = $0.9838$ corresponding to a HQBS with
a binding energy approximately 0.0014.

More importantly, when $V=0.2$, a RS peak is found in the
conductance around $E_{\rm RS}$ = $0.985$ caused by multiple
scattering between the two fingers leading to a resonant
transmission. The corresponding energy $E_{\rm RS}(n)$ can be
estimated using an infinite quantum well model with a zero point
energy measured from the subband top of the lower spin branch
$E^-_{0{\rm T}}$.
\begin{equation}
 E_{\rm RS}(n) = E^-_{0{\rm T}}
 + \left( \frac{n\pi}{L}\right)^2 \, ,
 \label{E_RS}
\end{equation}
in which $E^-_{0{\rm T}}$ = $0.9824$, and hence we obtain $E_R(2)
\sim 0.986$.

The probability densities shown in Figure \ref{fig5}(b)-(c) ($V =
\pm 0.2$) are typical for resonances formed between finger gates.
They are localized in the gate region, and due to the considerable
length of the system ($L=100$) the interference caused by the
fingers is well visible as oscillations of the probabilities
densities.

\begin{figure}[t]
\includegraphics[width = 0.48 \textwidth,angle=0] {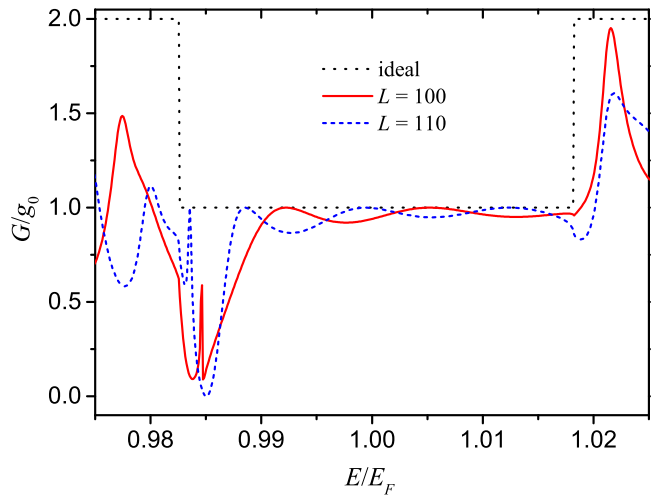}
\caption{
 (Color online) Conductance as a function of energy for a
 DFG system with $L$ = 100 (solid) and 110 (dash). $\alpha = 0.2$,
 $\beta = 0.1$, and $gB = 0.02$.
 }
 \label{fig6}
\end{figure}

In order to provide further evidence that the peak at the energy $E
= 0.985$, for gate potential $V = 0.2$ and distance $L = 100$
between the fingers (shown by red solid line in \fig{fig5}(a)) is a
RS caused by multiple scattering we test its length dependence. We
compare with results for $L = 110$ shown by a dashed blue curve in
\fig{fig6}.  The RS peak in the case of $L = 110$ is at $E = 0.984$
that is lower than for $L =100$ and in accordance with what is to be
expected. The dependence on the distance between the two fingers in
the DFG system allows us to identify the sharp peaks in the
conductance as a RS feature.


\section{Concluding Remarks}

In conclusion, We have developed a model to investigate the
interplay of the strong SO coupling and the Zeeman effect, in which
the lower spin branch contains a local band top in reciprocal space
forming a SO-Z gap. We have demonstrated that this particular
subband structure in SFG and DFG systems leads to interesting
spin-resolved electronic transport properties.

In the absence of the Dresselhaus effect, the spin-split subband
structure is symmetric with respect to the wave vector resulting in
a degeneracy of the subband threshold.  In this case, we identify
the physical mechanisms responsible for the appearance for
conductance mirror effect between the HQBS and EQBS in a SFG system.
However, in a DFG system, the HQBS caused by a positive FG potential
is strongly suppressed, but the EQBS feature remains significant.

In the presence of the Dresselhaus interaction, the subband
structure becomes asymmetric with respect to the wave vector. We
successfully predict the binding energy of the real BS in the lower
spin branch for both the SFG and the DFG systems.  Especially, a RS
can be found in a DFG system that is localized in the finger region
due to multiple scattering. Our theoretical prediction of a
formation of a BS and an RS mechanisms gives a hint for a design of
a SO-Z based spin electronic device.


 \begin{acknowledgments}
This work was supported by the MOST in Taiwan through Contract No.\
103-2112-M-239-001-MY3, the Icelandic Research and Instruments
Funds, and the Research Fund of the University of Iceland.
 \end{acknowledgments}


\end{document}